\documentclass{article}
\usepackage{graphicx}
\title{Notes On The Klein-Gordon Equation}
\author{Fredrick Michael* \\  \it{written October 2003}}
\begin{document}
\maketitle

\abstract{In this article, we derive the scalar Klein-Gordon equation 
from the formal information theory framework. The least biased probability distribution is obtained, and the scalar equation is recast in terms of a 
Fokker-Planck equation in terms of the imaginary time, or a Schroedinger equation for the proper time.
This method yields the Green's function parametrized 
by the proper-time, and is related to the results of Schwinger and Dewitt. 
The derivation can then allow the use of potentials as constraints along 
with the Hamiltonian or moments of the evolution. The information the-
oretic, analogously the maximum entropy method, also allows one to ex-
amine the possibility of utilizing generalized and non-extensive statistics
 in the derivation. This approach yields non-linear evolution in the Klein-
 Gordon-like partial differential equations. Furthermore, we examine the 
Klein-Gordon equation in curved space-time, and we compare our results 
to results obtained from path integral approaches.
}

\section{Introduction }
In this article, we derive the scalar Klein-Gordon equation from within the information theory (or maximum entropy) framework.  The least biased probability
distribution is obtained from the maximization of the entropy with suitable con-
straints.   The probability then  evolves under the scalar  wave equation with  a
parametrized mass term, the Klein-Gordon equation parametrized by the proper
time. The scalar equation is recast as  a Fokker-Planck equation in terms of the
parametrization  of imaginary time,  or a Schroedinger  equation for the proper
time. This method is akin to the Feynman parametrization of the mass term to
solve the Klein-Gordon wave equation and the derivation of the photon Green's
function, as well as the Schwinger-Dewitt proper-time formalism. The derivation
also allows one to use  a higher dimensional version of the methods of quantum
mechanics, specifically, the use of potentials in  a linear superposition with  the
Hamiltonian of the evolution via the use of observable constraints in the entropy
maximization procedure.  We then use  the method to examine  the underlying
stochastic differential equations and the microscopic evolution. We explore the
solution to  the wave equation  for a  particle moving in  curved spacetime  and
given the moments obtained from the underlying stochastic  microsocpic evolution, and obtain the  short-time transitional (conditional) probability  from the
maximum entropy variational principle.

In the past,  the proper-time  formalism has gained  acceptance in  the treatment of the relativistic wave equation both in 
flat and curved space-time.  The
method simply  involves a parametrization  of the  mass term,  and recasts  the
Klein-Gordon equation into  a partial differential  equation (PDE) with  an extra dimension (here) being that of the parametrization (the proper  time). The
solution  is then  obtained by  transforming from  the proper-time  to the  mass
term \cite{2,3}.  Also, well known  analytic continuation methods allow us  to recast
the result in terms of a higher dimension Schroedinger equation with the subsequent connection to known methods of low energy quantum  mechanics and the
Schwinger-Dewitt proper time  formalism.  What has not been done is to consider the resultant partial differential equation as a diffusion-like equation in the
context of probability theory and the maximum entropy method \cite{5,7}, or equivalently from within  information theory \cite{1}.  A mathematical reason for this is that the equations of the wave equation and Klein-Gordon form of equations are elliptical equations, and maximum entropy and information theory derive PDEs that are parabolic, of the diffusion, Fokker-Planck and Schroedinger forms, and that have 'friction'-like terms that allow for equilibriation with a heat bath in the thermodynamics perspective. 

 This interpretation of the relativistic elliptical wave equation and Klein-Gordon equations as parametrized higher dimensional parabolic equations following the work of Feynman will allow us to recast the problem as one of maximum entropy and information theory. From derived least biased distributions and their evolution parabolic PDEs we will obtain an equivalent description of the problem in terms
of the underlying stochastic differential equations. For the curved-spacetime case this will mean two new formulations, an extensive statistics and the recently developed nonextensive statistics formulation of the stochastic gravity approach. The moments obtained from
the stochastic description can then be used to obtain the short-time transition
probability for the evolution. The Klein-Gordon partial differential equation is
\begin{equation}
\includegraphics[width=90mm]{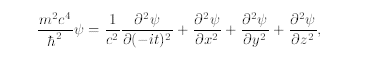}
\label{eqn1}
\end{equation}

and we rewrite the imaginary time as $-it=\tau$ . We can obtain this equation from the maximum entropy method, or equivalently the information theoretic framework as the information measure is equivalent to the entropy with appropriate constants of proportionality relating the bits of information to the measure of disorder the entropy measured by the Boltzmann constant. We begin the derivation by stating that we know the moments of the variables $(x,y,z,\tau)$ and maximize the Gibbs-Boltzmann entropy $<S>=-\int{PlnP}$, equivalently minimizing the Shannon information measure, given the following constraints parametrized by $\lambda$ 

\begin{equation}
\includegraphics[width=70mm]{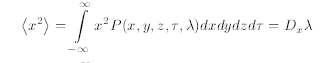}
\label{eqn2}
\end{equation}

\begin{equation}
\includegraphics[width=70mm]{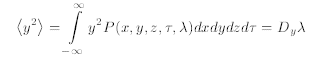}
\label{eqn3}
\end{equation}

\begin{equation}
\includegraphics[width=70mm]{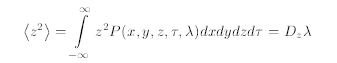}
\label{eqn4}
\end{equation}

\begin{equation}
\includegraphics[width=70mm]{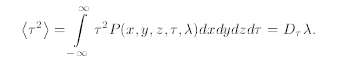}
\label{eqn5}
\end{equation}

Maximizing the entropy we obtain the least biased probability distribution function (PDF) normalized with N
\begin{equation}
\includegraphics[width=90mm]{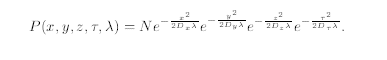}
\label{eqn6}
\end{equation}

This PDF solves the diffusion-like partial differential equation PDE
\begin{equation}
\includegraphics[width=100mm]{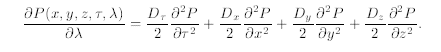}
\label{eqn7}
\end{equation}

We can rewrite the PDF via a Laplace or alternatively Fourier transform as 
\begin{equation}
\includegraphics[width=40mm]{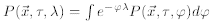}
\label{eqn8}
\end{equation}

and substitute in the PDE Eq. (\ref{eqn1}) and  obtain 
\begin{equation}
\includegraphics[width=100mm]{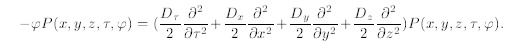}
\label{eqn9}
\end{equation}

With  the  identification of  the  diffusion   constants  as $D_{x,y,z}= 2$ and $D_\tau= 2{c^{-2}}$
and $\phi=-\frac{{m^2}{c^4}}{{\hbar ^2}}$,   we   recover  the   Klein-Gordon   equation    as   in   Eq.(\ref{eqn1}).    We   can
also   obtain   a   5D  Schroedinger   equation   using   this   method.    The   PDE   is   recast
in this  form if  we take the  time-like parameter and  Wick rotate  about the
imaginary axis to $\lambda=-i\alpha$. The resulting PDE is Schroedinger-like
\begin{equation}
\includegraphics[width=100mm]{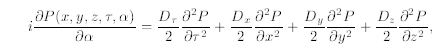}
\label{eqn10}
\end{equation}
and  we  can  identify  the   diffusion  constants  as  relating  the   'mass'  to  the  variance as $D_
{x,y,z} =-\frac{\hbar}{m^*}=\sigma^2$ and $D_\tau =-\frac{\hbar}{c^2 m^*}$.    Alternatively, one   can  work  with
the   Hamiltonian   directly.     From  the   Legendre   transform    of   the   relativistic   Lagrangian, one can  obtain the generalized  canonical momenta, and write  a relativistic Hamiltonian and  any potentials as the observables to be used as constraints in  the entropy  maxmization procedure.  Thus  the constraints will  be
the $< H >$, and  any potentials to  be included  such as external  potentials or
interactions $< V >$, and the entropy $< S >$ is maximized to  yield the least biased probability. In the  case of $< V >= 0$  and 
flat  space-time, the least biased
probability will solve the massive Klein-Gordon equation as in  Eq.(\ref{eqn1}).

\section{Connection with nonextensive statistics}
Having obtained a maximum entropy derivation of the KG type equation, a
question arises  as to the  form of  the equations obtained  if one were  to use  a
generalized entropy. Instead of the  Gibbs-Boltzmann entropy which yields the
extensive statistics,  one can use a  generalized form of  the entropy, the nonextensive entropy, which has become well known after the work  of C.  Tsallis \cite{5,7} and which  yields
non-linear partial differential equations for the evolution of the distribution.  In
cosmology, the non-extensive statistics  have been shown to  model 'small' systems  such as galactic matter  distributions, where the  range of the  interaction
(re: gravitational) is on the same scale as the system size \cite{10}, the solar neutrino
problem \cite{9} and systems that exhibit complex and non-linear dynamics \cite{11}.  The
present derived  result should  prove of  interest as  a new nonlinear theory  as
nonlinear Klein-Gordon models are of importance in high energy particle physics.
The non-extensive statistical entropy, or analogously the incomplete information
theoretic measure \cite{1} is
\begin{equation}
\includegraphics[width=60mm]{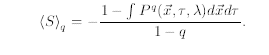}
\label{eqn11}
\end{equation}
This form of entropy is known to yield power law distributions. Maximizing the entropy with the observable moments as constraints yields the least biased power law distribution,
\begin{equation}
\includegraphics[width=80mm]{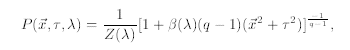}
\label{eqn15}
\end{equation}

where $ Z^2 (\lambda)\beta (\lambda)=const.$ and $Z(\lambda)$ is  the partition  function and is  related
to the  normalization as usual.   This distribution  solves the non-linear  partial
differential equation
\begin{equation}
\includegraphics[width=90mm]{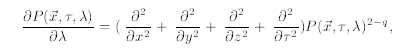}
\label{eqn16}
\end{equation}
where   we  have  set  $\hbar=c= 1$  and absorbed  any constants into  the variables
$(x, y,z,\tau)$.  We can  obtain the solution of the Klein-Gordon-like equation  if we
Laplace or alternatively Fourier transform according to
\begin{equation}
\includegraphics[width=60mm]{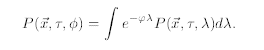}
\label{eqn17}
\end{equation}
Some comments should be made regarding Eq.(\ref{eqn16}). The nonlinear form of the PDE Eq.(\ref{eqn16}) is one that by q-parametrization is a controllable deviation from the Gaussian solution of $q->1$ towards the power law solution of Eq.(\ref{eqn15}). The nonlinear terms of the traditional nonlinear Klein-Gordon equations of particle high energy physics are solved by functions that are deviations from the Gaussian, squeezed or stretched Gaussians. These nonlinear deviations are present by parametrized nonlinearity $q$ and the PDE equation Eq.(\ref{eqn16}) should be considered as a new class of nonlinear Klein-Gordon equations.
Another comment is to be made regarding curved space-time applications of this equation Eq.(\ref{eqn16}). The curvature terms are included as the diffusion coefficients of the PDE Eq.(\ref{eqn4}) as the curvature tensor terms are for diagonal tensors or tensors that can be transformed to diagonal tensors the inverses of the diffusion coefficients. These curvature terms cause the PDE Eq.(\ref{eqn4}) to become nonlinear and the solutions of which become deviations from the Gaussian as discussed. 
Therefore the possibility exists here that the complicated nonlinearities due to nonlinear terms included to describe high energy particles and nonlinear terms introduced to describe curved space-time can be described or modeled by the $q$-parametrized nonlinearity of Eq.(\ref{eqn16}) which has known solutions of the power-law form of Eq.(\ref{eqn15}). Also it is important to note that a recent advance in nonextensive quantum mechanics \cite{12} is promising for a relativistic Schroedinger-like PDE approach to nonextensive relativistic quantum mechanics. This will be pursued in subsequent work.   

\section{Stochastics}

The diffusion equation Eq.(\ref{eqn4}) is a Fokker-Planck equation.  This macroscopic
evolution equation for the PDF allows us to obtain the underlying Ito-Langevin stochastic differential equations \cite{4}.  The individual trajectories are
\begin{equation}
\includegraphics[width=70mm]{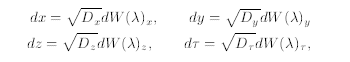}
\label{eqn18}
\end{equation}
where the $dW_{x,y,z,\tau}$ are Wiener  processes,  and  are composed of  Gaussian  white
noise with delta  correlations.  The case of  the non-extensive statistics  derived Klein-Gordon-like PDE Eq.(\ref{eqn16}) is somewhat more complicated.  The underlying stochastic evolution is of  the statistical feedback form \cite{6}, and from Eq. (\ref{eqn16}) we obtain
\begin{equation}
\includegraphics[width=90mm]{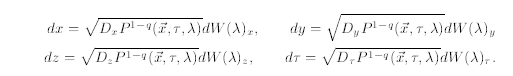}
\label{eqn19}
\end{equation}
We note  again that the macroscopic  distribution PDF appears in the  underlying
microscopic stochastic evolution.  This statistical feedback 'loop' between the macroscopic evolution of probability distribution $and$ the microscopic stochastic trajectory(s) can be  viewed
as the source of the nonlinearity in the  PDE from this perspective.  The modification of the trajectories via coupling to the large scale macroscopic evolution, alternatively the memory effect or the historical trends
squeezes the PDF from the usual Gaussian type into a power-law, and can be a
prototype for nonlinear deviations from the usual parametrized Klein-Gordon based Gaussian
statistics.  We note again that due to the presence of curvature, metrics with nonlinear terms  will contribute terms  to the Klein-Gordon  equation that will  cause
deviations from a Gaussian \cite{2} and that can be modeled by the $q$-parametrized nonlinearity.  We will examine the regular form of this evolution in the context of curved spacetime in the next section. Mapping the regular curved space-time parametrized evolution to the nonextensive statistics $q$-parametrized evolution we leave for subsequent work.

\section{Klein-Gordon Equation in Curved Spacetime}
The Klein-Gordon equation in curved spacetime from within  the proper-time
formalism of Schwinger and Dewitt has been treated, at least from the Gaussian
approximation, by Bekenstein \cite{2,3}.  The  equation is derivable from  an action
principle
\begin{equation}
\includegraphics[width=80mm]{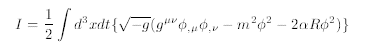}
\label{eqn20}
\end{equation}
The  Euler-Lagrange  equation  then  yields  the  Klein-Gordon  equation  in curved
space-time, with $c= \hbar= 1$

\begin{equation}
\includegraphics[width=50mm]{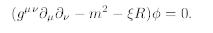}
\label{eqn20b}
\end{equation}

We  will  solve  the  general  case  first,  with $R=R(\vec x)$. The solution method  will
depend on  the use  of stochastic calculus  and the  maximum entropy method.
We define  the two-point function $P(\vec x,\tau, \lambda;\vec x',\tau', \lambda')$.  This  function evolves 
in Eq. (\ref{eqn20b}).  Writing generally, the backwards PDE is
\begin{equation}
\includegraphics[width=80mm]{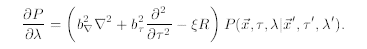}
\label{eqn21}
\end{equation}

We can transform  this  equation  into a  standard  form if  we  transform according
to

\begin{equation}
\includegraphics[width=90mm]{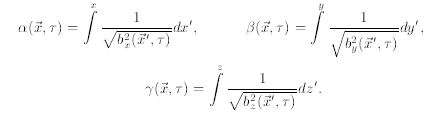}
\label{eqn22}
\end{equation}
The transformed equation is
\begin{equation}
\includegraphics[width=80mm]{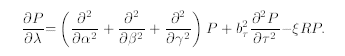}
\label{eqn23}
\end{equation}
%The transformed equation is

We  redefine  the  function $P$ as $P(\vec \alpha, \tau, \lambda)=\pi e^{-U(\vec \alpha)}$. Upon substitution  and gathering terms, we have
\begin{equation}
\includegraphics[width=90mm]{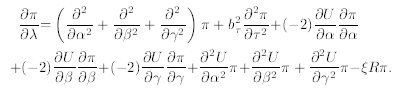}
\label{eqn24}
\end{equation}

The curvature term is eliminated if we make the identification of the first term with $\xi R$. Upon replacing the original  arguments, we have

\begin{equation}
\includegraphics[width=90mm]{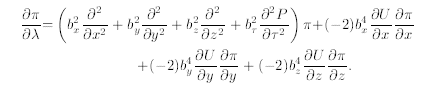}
\label{eqn25}
\end{equation}
%\begin{equation}
%\includegraphics[width=90mm]{eqn26.png}
%\label{eqn26}
%\end{equation}
This equation is now in the backwards Chapman-Kolmogorov form. The two-point function, or the conditional distribution, is a solution of this
equation, as it is a solution of the forward Chapman-Kolmogorov equation, the
Fokker-Planck equation,
\begin{equation}
\includegraphics[width=90mm]{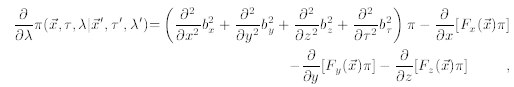}
\label{eqn27}
\end{equation}
where we have renamed the  drift coefficients as ${F_x}(\vec x)= 2{b^4}_x {\frac{\partial U}{\partial x}}$,  and  similarly for  the other coefficients and have dropped the prime superscripts of the variables.  The  underlying  stochastic differential equations  are
\begin{equation}
\includegraphics[width=110mm]{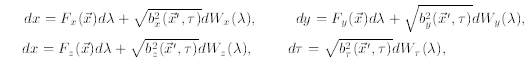}
\label{eqn28}
\end{equation}
where the $dW$ are Wiener  processes and are delta correlated as before. The
time averages of the descretized stochastic processes are $(\Delta x=x-{x_o})$
\begin{equation}
\includegraphics[width=70mm]{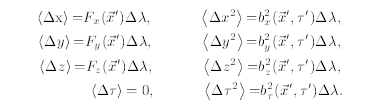}
\label{eqn29}
\end{equation}

These averages coincide with conditional expectation values such as
\begin{equation}
\includegraphics[width=70mm]{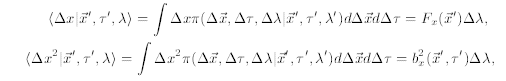}
\label{eqn30}
\end{equation}
and so on for the other coordinate conditional expectation values. We solve for
the short-time conditional (transition) probability distribution as follows. The
conditional variances about the means are now the constraints  as
we maximize the conditional entropy
\begin{equation}
\includegraphics[width=100mm]{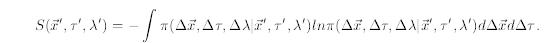}
\label{eqn31}
\end{equation}
Explicitly we maximize the following expression
\begin{equation}
\includegraphics[width=100mm]{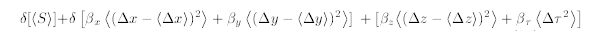}
\label{eqn32}
\end{equation}
The maximization yields the least biased probability
\begin{equation}
\includegraphics[width=100mm]{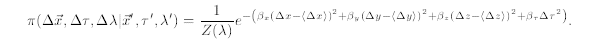}
\label{eqn33}
\end{equation}
The Lagrange multipliers are obtained from the partition function via functional
relationships between the multipliers and the moments.  As an example, the x-coordinate multipliers are
\begin{equation}
\includegraphics[width=30mm]{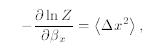}
\label{eqn34}
\end{equation}
and the partition function is $Z=\int \pi$.  Upon integration, this yields
\begin{equation}
\includegraphics[width=50mm]{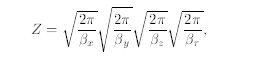}
\label{eqn35}
\end{equation}
and we can solve for the multipliers using the relationships Eq.(\ref{eqn34}) to obtain
\begin{equation}
\includegraphics[width=60mm]{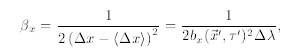}
\label{eqn36}
\end{equation}
and similarly  for   the   multipliers  in   the   other   coordinates,  in  terms   of   their
moments. Gathering the  terms together and substituting for the moments, we
have
\begin{equation}
\includegraphics[width=110mm]{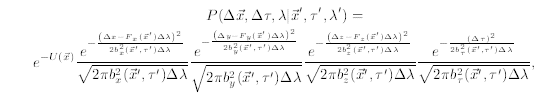}
\label{eqn36b}
\end{equation}

We can compare this result  with the proper-time  formalism of Schwinger-Dewitt and  the results of Bekenstein \cite{2,3} if  we wick rotate to the proper time
$\lambda=- i s$.  The Green's function from the path  integral approach is
\begin{equation}
\includegraphics[width=90mm]{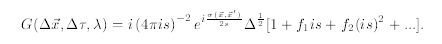}
\label{eqn37}
\end{equation}
The factors  in   the  expression   are $\sigma(\vec x;\vec x')=\sigma(\vec x';\vec x)$ and stands for half the proper distance squared $d{S^2}=g_{\mu \nu} dx_\mu dx_\nu $,  or minus  half  the  proper time
squared,  and   contains  the  metric  coefficients,  which  we  have re-written  as  inverses and are contained in the diffusion coefficients ${b^2}_{x,y,z,\tau}$ .    The  quantity  $\Delta$ is $\Delta(\vec x;\vec x')=-g(\vec x)^{\frac{1}{2}} D_{VM} g(\vec x')^{\frac{1}{2}} $, and $D_{VM}$ is the VanVleck-Morreta determinant $D_{VM}=Det(- \frac{\partial^2 \sigma}{\partial x^{\mu} \partial x^{\nu}})$. The flat space-time result is $G(\Delta \vec x, \Delta \tau, \lambda)_o =i (4 \pi i s)^{-2} e^{i \frac{\pi_o (\vec x;\vec x')}{2 \sigma}}  \Delta^{\frac{1}{2}}$, and Bekenstein solves the Gaussian case for the geodesic extreme paths and obtains
\begin{equation}
\includegraphics[width=90mm]{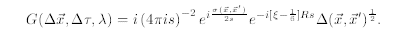}
\label{eqn38}
\end{equation}
We note again that the diffusion  coefficients  are   the   inverse  of  the   metric  coefficients and are here included in the factor $\sigma$. The normalizations are expectedly different, as is the $R$ curvature scalar term, as the solution in the present case
takes into account terms higher in the expansion.  The short-time transition
probability is accurate to order $\theta(\Delta \lambda)$. We regain the flat space-time case where
$R=0$ if we let the means vanish (the drift coefficients), as the scalar  curvature
is included in the $x,y,z,\tau$ coordinates' means, and assume constant (unit) diffusion
coefficients.

\pagebreak
\section{Conclusion }

In this article we have derived the scalar Klein-Gordon equation  from within
themaximumentropymethod,obtaining the (Wick rotated) proper-time analog and 
solutions.  We have also generalized  the approach to include other than
Gibbs-Boltzmann entropies, namely  the non-extensive statistics  entropy of C.
Tsallis.  This  resulted in a  new nonlinear form  of the Klein-Gordon  equation,
with its nonlinear proper time PDE. We then examine the Klein-Gordon equa-
tion  in curved space-time, and obtain a general  solution in which the diffusion
coeffients  are related  to the inverse metric coefficients.  Also, the curvature
scalar is then included in the drift.  We compare our result with the  results of
Bekenstein and Dewitt.  We leave the application to specific curved space-time
metrics to future work.

\pagebreak
 * Email address Fredrick Michael, PhD. PI Michael Research R$\&$D, fnmfnm2@gmail.com, fmicha3@uic.edu .

\end{document}